\input{epsf}
\documentstyle[aps,preprint]{revtex}
\begin{document}
\draft
\tightenlines
  
\title{Asymptotic behavior of the order parameter in
a stochastic sandpile}
\author{Ronaldo Vidigal\footnote{E-mail: rvidigal@dedalus.lcc.ufmg.br} 
and Ronald Dickman\footnote{E-mail: dickman@fisica.ufmg.br}}
\address{
Departamento de F\'{\i}sica, ICEx, 
Universidade Federal de Minas Gerais,\\ 
30123-970 
Belo Horizonte - Minas Gerais, Brazil}

\date{\today}
\maketitle 

\begin{abstract} 

We derive the first four terms in a series for the
order paramater (the stationary activity density $\rho$)
in the supercritical regime of a one-dimensional
stochastic sandpile; in the two-dimensional case
the first three terms are reported.
We reorganize the pertubation theory for the model, 
recently derived using a path-integral formalism 
[R. Dickman e R. Vidigal, J. Phys. A 
{\bf35}, 7269 (2002)], to obtain an expansion for
stationary properties.
Since the process has a strictly conserved particle
density $p$, the Fourier mode $N^{-1} \psi_{k=0} \to p$, 
when $N \to \infty$,
and so is not a random variable.  Isolating this mode, we obtain a
new effective action leading to an expansion for $\rho$
in the parameter $\kappa \equiv 1/(1+4p)$. 
This requires enumeration and numerical evaluation of more than 200$\,$000
diagrams, for which task we develop a computational algorithm.  
Predictions derived from this series are in good accord with
simulation results.  We also
discuss the nature of correlation functions and one-site 
reduced densities in the small-$\kappa$ (large-$p$) limit.

\end{abstract} 
\vspace{1cm} 

\newpage
\section{Introduction}
\newcommand{\be}{\begin {equation}}
\newcommand{\ee}{\end {equation}}
\newcommand{\bea} {\begin {eqnarray}}
\newcommand{\eea} {\end {eqnarray}}
\newcommand{\re}{\mathrm{I}\!\mathrm {R}}

Sandpiles are the principal examples of self-organized criticality (SOC)
\cite {BTW,DD,GG,SJD,VZ1}.
Sandpiles with a strictly conserved particle density (so-called
{\it fixed-energy sandpiles} or FES 
\cite {DVZ}),  exhibit an
absorbing-state phase transition \cite{MD,HH,V}, 
rather than SOC, and have recently attracted much interest.
Until now, most quantitative results for FES have been 
based on simulations \cite{VDM,RPV,DAM,RTO,lubeck,heger}, 
an important exception being the solution by Priezzhev {\it et al.}
\cite{PIP} of a directed, fixed-energy version of the Maslov-Zhang
model \cite{MZ}, via the Bethe ansatz.
Recently, a time-dependent perturbation theory based on the 
path-integral formalism was derived for a stochastic sandpile \cite{DV1}.
In \cite{RD1} the series expansion for the one-dimensional case was
extended using operator methods.  

In the present work, the perturbation theory developed in 
\cite{DV1} will be reformulated, leading to an expansion for stationary
($t \to \infty$) properties instead of the short-time expansion
obtained previously.  The expansion parameter is $\kappa \equiv 1/(1+4p)$,
where $p$ denotes the particle density, a conserved quantity.

Our analysis depends on two basic tools. 
One is an operator formalism for Markov processes,
of the kind developed by Doi \cite{DOI}, 
and which has been applied to various models exhibiting
nonequilibrium phase transitions \cite{RD4,DJ,ID,DWJ,ZRP}. 
The second is an exact mapping, devised by
Peliti, of a  Markov process to a path-integral representation
\cite{PLT,DV2}.
This approach is frequently used to generate the effective action
corresponding to a process, for subsequent analysis via
renormalization group (RG) techniques.  In the present instance our
immediate objective is not a RG analysis but an expansion for
the order parameter.
In the path-integral formalism the probability generating function
is written in terms of functional integrals over the fields $\psi (x,t)$
(whose expectation is the particle density at site $x$), and 
an auxiliary field $\tilde{\psi}(x,t)$.
Our reformulation of the effective action is based on the observation that,
due to particle conservation, the Fourier mode
 $N^{-1}\psi_{k=0}$ is not a random variable, but rather has the fixed value
$p$ when $N$, the number of lattice sites, goes to infinity.

We consider Manna's stochastic sandpile in its fixed-energy 
(particle-conserving) version
\cite{DAM,DV1,MS,MS1}. 
The configuration is specified by the occupation number
$n$ at each site; sites with $n \geq 2$ are said to be 
{\it active}, and have a positive rate of
{\it toppling}.  When a site topples, it loses exactly two
particles (``grains of sand"), which move randomly and 
independently to nearest-neighbor (NN) sites.
(Any configuration devoid of active sites is {\it absorbing},
i.e., no futher evolution of the system is possible once such
a configuration is reached.)
In this work, as in 
\cite{DV1,RD1}, 
we adopt a toppling rate of
$n(n\!-\!1)$ at a site having $n$ particles, which leads us to define the
order parameter as $\rho = \langle n(n-1) \rangle$. 
While this choice of rate represents a slight departure from the
usual definition (in which all active sites have the same toppling rate), 
it leads to a much simpler
evolution operator, and should yield the same
scaling properties
\cite{DV1}. 
Preliminary simulation results 
\cite {RD3} 
indicate that in one dimension the model exhibits a continuous phase transition
at $p_c \!=\! 0.9493$.  
  
The balance of this article is organized as follows.
In Sec. 2, we discuss the reorganization of the action, and in Sec. 3
develop a perturbation expansion for the activity density in the
supercritical regime.  Sec. 4 presents the diagrammatic expansion rules and
the resulting expansion.  Predictions for the activity density are reported and
compared against simulation in Sec. 5, while in Sec. 6 we examine 
correlation functions and higher moments of the density.
In Sec. 7 we present a brief discussion
of our results.

\section{Effective action}

As shown in \cite{DV1}, the master equation for the stochastic 
sandpile can be written in the form
\be
\label{Eq:MST}
\frac{d\mid\Psi\rangle}{dt}=L\mid\Psi\rangle,
\ee
where
\be
\label{Eq:PSI}
\mid\Psi\rangle=\sum_{\{n_{i}\}}p(\{n_{i}\},t)\mid \{n_{i}\}\rangle,
\ee
where $p(\{n_{i}\},t)$ is the probability of the configuration having
occupation numbers $\{n_{i}\}$ and
$\mid \{n_{i}\}\rangle$ is the direct product of states
$\mid n_{j}\rangle$, 
representing exactly $n_{j}$ 
particle at site $j$.
In one dimension, the evolution
operator takes the form
\be
\label{Eq:DP}
L=\sum_{i}
\left [ \frac{1}{4}(\pi_{i-1}+\pi_{i+1})^{2}-\pi_{i}^{2}\right ]a_{i}^{2}
\equiv\sum_{i}L_{i}.
\ee
Here $a_{i}$ e $\pi_{i}$ are, 
respectively, destruction and creation 
operators associated with 
site $i$, defined via
\be
\label{Eq:AI}
a_{i}\mid n_{i}\rangle=n_{i}\mid n_{i}-1\rangle
\ee 
and
\be
\label{Eq:PI}
\pi_{i}\mid n_{i}\rangle=\mid n_{i}+1\rangle.
\ee

As shown in \cite{DV1}, the evolution
operator in Fourier representation is given by 
\be
\label{Eq:OE}
L=-N^{-3}\sum_{k_{1},k_{2},k_{3}}\omega_{k_{1},k_{2}}
\pi_{k_{1}}\pi_{k_{2}}a_{k_{3}}a_{-k_{1}-k_{2}-k_{3}},
\ee
with $\omega_{k_{1},k_{2}}=1-\cos k_{1}\cos k_{2}$; 
the sums are over the first
Brillouin zone.

As explained in \cite{DV1}, the expectation of
any observable $A(\{n_i\})$ can be written in terms of a functional integral 
\be
\label{Eq:FAM}
\langle {\cal A}\rangle=\int {\cal D}\tilde{\psi}
{\cal D}\psi{\cal A} \; {\cal G}[\psi,\tilde{\psi}],
\ee
where ${\cal A}(\psi,\tilde{\psi})$ is a function of the fields $\psi$ and
$\tilde{\psi}$ corresponding to observable $A$, and
\be
\label{Eq:GM}
{\cal G}[\psi,\tilde{\psi}]\equiv\exp\left[-N^{-1} \! \int_{0}^{t}dt^{'}\sum_{k} \tilde{\psi}_{k}  
\dot{\psi}_{-k}+\int_{0}^{t}dt^{'}{\cal L}_{I}\right]\equiv\exp\left[-N^{-1} \!  
\int_{0}^{t}dt^{'}{\cal L}_{0}+\int_{0}^{t}dt^{'}{\cal L}_{I}\right],
\ee
with the interaction given by
\bea
\label {Eq:Li}
{\cal L}_{I}& = & -N^{-3}  
\sum_{k_{1},k_{2},k_{3}}\omega_{k_{1},k_{2}}
\tilde{\psi}_{k_{1}}\tilde{\psi}_{k_{2}}\psi_{k_{3}}
\psi_{-k_{1}-k_{2}-k_{3}} \nonumber \\ 
 & - & 2 N^{-2}\sum_{k_{1},k_{2}}\omega_{k_{1},0}
 \tilde{\psi}_{k_{1}}\psi_{k_{2}}\psi_{-k_{1}-k_{2}}.      
\eea
The field $\psi$ is closely related to the occupation number \cite{DV2}.
In particular, the activity density is given by
\be
\label{Eq:AMG}
\rho (t) \equiv N^{-1}\sum_{j}\langle n_{j}(n_{j}- 1)\rangle
=N^{-2}\sum_{k}\langle\psi_{k}\psi_{-k}\rangle\;\;,
\ee
while for the particle density we have
\be
\label{Eq:PD}
\phi \equiv N^{-1}\sum_{j}\langle n_{j}\rangle
=N^{-1} \langle\psi_{k=0} \rangle\;\;.
\ee
In \cite{DV1} equations (\ref{Eq:FAM}) - (\ref{Eq:AMG}) 
serve as the starting point 
for a diagrammatic expansion of $\rho(t)$ in powers of time.  
We now show how 
these relations may instead be used as the basis for an expansion
of the stationary activity density
$\rho_\infty \equiv \lim_{t \to \infty} \rho(t)$.

In writing equations (\ref{Eq:GM}) and (\ref{Eq:Li}) we have assumed
a Poisson-product distribution, with expectation $p$, for
the initial occupation numbers $n_i$.  Thus $\langle\psi_{k=0} \rangle = Np$,
a constant of the motion, since the number of particles is conserved.
In the infinite-size limit, the law of large numbers implies
that $N^{-1} \psi_{k=0}= p$, 
and is no longer a random variable.
We may therefore isolate all terms with
$k=0$ in equation (\ref{Eq:Li}) setting each factor $N^{-1}\psi_{k=0}$ equal to 
$p$.  (Observe as well that $\tilde{\psi}_{k=0}$, the variable 
conjugate to $\psi_{k=0}$, is no longer needed.)  
As a result of this procedure
${\cal G}[\psi,\tilde{\psi}]$ assumes the form
\be
\label{Eq:G}
{\cal G}[\psi,\tilde{\psi}]\equiv\exp\left[-N^{-1} \int_{0}^{t}dt^{'}
\sum_{k\neq 0} (\tilde{\psi}_{k} 
\dot{\psi}_{-k}+\gamma_{k}\tilde{\psi}_{-k} \psi_{k})
+\int_{0}^{t}dt^{'}{\cal L}^{'}_{I}\right],
\ee
with
\be
\gamma_{k}=4\;p\;\omega_{-k,0}=4\;p\;(1-\cos k)
\label{gamma}
\ee
and the modified interaction
\bea
\label{Eq:LI}
{\cal L}^{'}_{I}& = & -N^{-3} \sum_{k_{1},k_{2},k_{3} \neq 0} 
\omega_{k_{3},-k_{1}-k_{2}-k_{3}}\psi_{k_{1}}\psi_{k_{2}}\tilde{\psi}_{k_{3}}
\tilde{\psi}_{-k_{1}-k_{2}-k_{3}} \nonumber \\
& - &  2pN^{-2}\sum_{k_{1},k_{3} \neq 0}
\omega_{k_{3},-k_{1}-k_{3}}\psi_{k_{1}}\tilde{\psi}_{k_{3}}
\tilde{\psi}_{-k_{1}-k_{3}}
 -  2N^{-2}  \sum_{k_{1},k_{2} \neq 0}
 \omega_{-k_{1}-k_{2},0}\psi_{k_{1}}\psi_{k_{2}}\tilde{\psi}_{-k_{1}-k_{2}}      
\\
&-& p^{2}N^{-1}\sum_{k_{3} \neq 0}\omega_{k_{3},-k_{3}}\tilde{\psi}_{k_{3}}
\tilde{\psi}_{-k_{3}} \nonumber \\ 
\eea
Here it is understood that none of the wavevectors associated with
the fields $\psi$ and $\tilde{\psi}$ may be zero.
The bilinear part of the action in equation (\ref{Eq:G}) represents independent
diffusion of particles at rate $4p$ \cite{DV2}.
The appearance of diffusion at rate $4p$ in ${\cal L}_0$ may be understood
intuitively as follows.  The rate of diffusion events at a given site is
$n(n-1)$, i.e., twice the number of distinct pairs, so that the diffusion rate
{\it per pair} is 2.  The diffusion rate per particle is the
twice the diffusion rate per pair times the number of pairs per 
particle, or $4(n-1) \simeq 4n \simeq 4p$ if $p \gg 1$.
Unlike the original representation of equation (\ref{Eq:GM}), the important
control parameter $p$ now appears explicitly in the action, rather
than being ``hidden" in the initial probability distribution.
It is worth noting that this reorganization of the action is not  
readily implemented in the operator representation,
equation (\ref{Eq:DP}), because in this case it is the {\it operator}
$N^{-1} \sum_i \pi_i a_i$ that assumes a fixed value $p$.

\section{Perturbation expansion}

Let equation (\ref{Eq:G}) with ${\cal L}^{'}_{I}\equiv 0$ define ${\cal G}_{0}$; 
equation (\ref{Eq:FAM}) with ${\cal G}_{0}$ in place of 
${\cal G}$ defines the free expectation $\langle {\cal A}\rangle_{0}$. 
Then for $k \neq 0 $ we have \cite{DV1}
\bea
\langle \psi_{k}(s)\rangle_{0} & = & 0 \nonumber \\
\langle \tilde{\psi}_{k}(s)\rangle_{0} & = & 0 
\eea
and the basic contraction or propagator is
\be
\label{Eq:ELC}
\langle \psi_{k^{'}}(u)\tilde{\psi}_{k}(s)\rangle_{0} =  
N\delta_{k^{'},-k}\Theta(u-s)e^{-\gamma_{k}(u-s)},
\ee
where $\Theta$ represents the step function.  As usual in this formalism,
$\Theta(0)=0$ \cite{DV2}. The free expectation
of $n$ fields $\tilde{\psi}$ and $n$ fields $\psi$ 
is given by the sum of all possible products of $n$ contractions.

The expectation of an observable can be written in the form 
\be
\langle {\cal A} \rangle = 
\left\langle {\cal A } e^{\int_0^t {\cal L}' dt'} \right\rangle_0
\label{Eq:A}
\ee
which can be expressed in terms of free expectations if we expand
the exponential. In this expansion, each field 
$\tilde{\psi}_{k}(\tau)$ must be contracted with a field 
$\psi_{-k}(\tau^{'})$, with $\tau^{'}>\tau$. 
At $n$-th order there is a factor of $1/n!$ and integrations
$\int dt_1 \cdots dt_n$ over the interval $[0,t]$.  We impose
the time ordering
$t\geq t_{1}\geq 
t_{2}\geq \ldots \geq t_{n}\geq 0$, thereby cancelling the factor $1/n!$.
We adopt a diagrammatic
notation \cite{DV1} in which fields
$\psi(\tilde{\psi})$ are represented by lines entering (leaving)
a vertex.  All lines are directed to the left, the direction of
increasing time.  The first term in ${\cal L}^{'}_{I}$, 
equation (\ref{Eq:LI}), corresponds to a vertex with four lines 
(``4-vertex"), 
the second and third to vertices with three lines (``3-vertex"), while the
fourth, with two lines exiting, will be referred to as a ``source."
figure 1 shows the vertices associated with ${\cal L}^{'}_{I}$, as well as
the ``sink" corresponding to the observable $\rho$.
Vertex b will be called a ``bifurcation" and c a``junction".
In this way, the activity density
\be
\label{Eq:RHO1}
\rho=N^{-2}\sum_{k}\langle\psi_{k}\psi_{-k}\rangle=N^{-2}\sum_{k}\langle\psi_{k}\psi_{-k}e^{\int_{0}^{t}  
dt^{'}{\cal L}^{'}_{I}}\rangle_{0}
\ee        
takes the form
\bea
\label{Eq:RHO2}
\rho & = & 
+ N^{-2}  
\langle 
\psi_{k=0}^2 e^{\int_{0}^{t}dt^{'}{\cal L}^{'}_{I}}
\rangle_{0} 
+ N^{-2}\sum_{k \neq 0}\langle \psi_k \psi_{-k}
e^{\int_{0}^{t}dt^{'}{\cal L}^{'}_{I}}
\rangle_{0}
\nonumber \\
& = & 
p^{2}+N^{-2}\sum_{k\neq  
0}\langle\psi_{k}\psi_{-k}e^{\int_{0}^{t}dt^{'}{\cal L}^{'}_{I}}\rangle_{0}.
\eea  

Consider the first order term.  From figure 1 it is evident that the only
vertex that can be contracted with the sink (without leaving dangling
lines) is the source.  This {\it simple loop}, shown as the first diagram
on the right hand side of figure 2, makes the contribution
\bea
\label{Eq:D1}
& - & 2p^{2}N^{-1}\sum_{k}\omega_{k,-k}\int_{0}^{t}e^{-2\gamma_{k}(t-t_{1})} \nonumber \\
& = & \frac{p}{4}\int_{-\pi}^{\pi}\frac{dk}{2\pi}(1+\cos k)[e^{8p(1-\cos k)t}-1] \nonumber \\
& = & \frac{p}{4}\left\{e^{-8pt}[I_{0}(8pt)+I_{1}(8pt)]-1\right\}.
\eea
where the prefactor 2 is a combinatorial factor and 
$I_{\nu}$ denotes the modified Bessel function. 
Here we used
\be
N^{-1} \sum_k \stackrel {N \to \infty} {\longrightarrow}
\int_{-\pi}^{\pi}\frac{dk}{2\pi}  \;.
\label{ident}
\ee
Thus this diagram yields the contribution identified in Ref. \cite{DV1} as
$\rho_{max}(t)$, the sum of all contributions at order $n = 1, 2, 3,...$
proportional to $p^{n+1}$, the highest power of $p$ allowed at a given order.
In the limit $t \to \infty$ the contribution to the activity from this
term is $-p/4$.

To study the stationary regime it is convenient to use the
Laplace transform.  For example, the Laplace transform of the
contribution due to the simple loop, equation (\ref{Eq:D1}), is
\be
\label{Eq:TLD1}
-\frac{2p^{2}}{s}\int_{-\pi}^{\pi}
\frac{dk}{2\pi}\frac{1-\cos^{2}k}{s+8p(1-\cos k)}
\ee
where $s$ denotes the transform variable.
Using the property $\lim_{t \to \infty} f(t) = \lim_{s \to 0} s \tilde{f}(s)$,
we obtain the limiting contribution $-p/4$ directly.

Consider an arbitrary diagram $D$ of $n$ vertices, and denote 
the time-dependent factors in its
contribution to $\rho (t)$ by $f_D(t)$.  The Laplace transform of
this contribution has the form
\bea
\label{Eq:LA}
\tilde{f}_D(s) & = & \int_{0}^{\infty}dt\;e^{-st}\int_{0}^{t}dt_{1}\int_{0}^{t_{1}}dt_{2}\ldots  
\int_{0}^{t_{n-1}}dt_{n}e^{-\alpha_{1}(t-t_{1})-\alpha_{2}(t_{1}-t_{2})\ldots-\alpha_{n}(t_{n-1}-t_{n})}
\nonumber \\
& = &  
\int_{t_{1}}^{\infty}dt\int_{t_{2}}^{\infty}dt_{1}\ldots\int_{0}^{\infty}dt_{n}e^{-(\alpha_{1}+s)(t-t_{1}) 
-(\alpha_{2}+s)(t_{1}-t_{2})\ldots(\alpha_{n}+s)(t_{n-1}-t_{n})-st_{n}}\nonumber \\
& = & [s(\alpha_{1}+s)(\alpha_{2}+s)\ldots(\alpha_{n}+s)]^{-1},
\eea 
where the $\alpha_i$ are functions of the wavevectors.  Then we have
\be
\label{Eq:PLA}
\overline{f}_D \equiv \lim_{t \rightarrow \infty} f_D(t)
=\prod_{i=1}^{n}\frac{1}{\alpha_{i}}.
\ee

The factors $\alpha_i$ may be determined via the following procedure.
Draw the diagram with all vertices in order, and draw vertical lines
through each vertex.  Then $\alpha_i$ is the sum of the factors
$\gamma_q$ for all propagators between the vertical lines associated
with vertices $i$ and $i-1$ (here $t=0 \equiv t_0$), {\it regardless
of whether or not these propagators link vertices} $i$ and $i-1$.

For example, a diagram composed of $n$ simple loops (see figure 2)
makes a contribution of
\be
\label{Eq:D1N}
\frac{(-1)^{n}2^{n}p^{2}}{s}\left[\int_{-\pi}^{\pi}\frac{dk}{2\pi}
\;\frac{1-\cos^{2} k}{s+8p(1-\cos k)}\right]^{n},
\ee
to $\tilde{\rho}(s)$, and so its contribution to 
$\rho_\infty$ is
\be
\label{Eq:D1NE}
\frac{(-1)^{n}2^{n}p^{2}}{(8p)^{n}}=(-1)^{n}p^{2}\frac{1}{(4p)^{n}}.
\ee
Summing on $n$, we find the contribution due to this sequence of
diagrams to the reduced activity
$\bar{\rho} \equiv \lim_{t\rightarrow \infty}(\rho/p^{2})$:
\be
\label{Eq:RHOB}
\sum_{n=1}^{\infty}\left(\frac{-1}{4p}\right)^{n}=-\frac{1}{1+4p} 
\equiv  -\kappa \;.
\ee
In certain cases it is straightforward to replace a simple loop with the 
infinite
sum of 1, 2, 3, ... loops.  This procedure, illustrated graphically
in figure 2, will be called {\it dressing} a loop.

Figure 3 shows a three-vertex diagram not included in
the sequence equation (\ref{Eq:RHOB}).  It
makes the following contribution to
$\bar{\rho}$:
\be
\label{Eq:D3}
\frac{-32p}{4p(8p)^{2}}\int_{-\pi}^{\pi}\frac{dk}{2\pi}\; (1+\cos k)
\int_{-\pi}^{\pi} \frac{dq}{2\pi}\;               
\frac{1-\cos q\cos (k-q)}{3-\cos q -\cos k -\cos (k-q)}\;\;\;.
\ee
The integral over wavevector $q$ arises frequently in the diagrammatic
series and can be evaluated in closed form:
\begin{eqnarray}
\label{Eq:INTK}
\nonumber
I(k) &=&\int_{-\pi}^{\pi}\frac{dq}{2\pi}\frac{1-\cos q\cos (k-q)}{3-\cos q -\cos k -\cos (k-q)}.
\\
&=&  \frac{1}{2} \left[ \frac{3 - c - \sqrt{(1-c)(7-c)}}{1+c}
+ \sqrt{\frac{1-c}{7-c}} \right]
\end{eqnarray}
where $c$ denotes $\cos k$.

In any diagram (beyond the set included in figure 2), 
we may insert any number of loops immediately to the right
of the sink.  That is, the sink may be replaced by a dressed loop.
The same applies to the rightmost source, vertex $n$.  
The result is that the contribution of the original
diagram is multiplied by $[4p/(1+4p)]^2$.  Once this factor is included,
no diagram with a 4-vertex immediately to the left of the rightmost
source (i.e., in position $n-1$) or immediately to the right of
the sink (position 1) need be included in the series.

\section{Diagrammatic Analysis}

To begin we define the rules for constructing diagrams in the series for
$\overline{\rho}$ \cite{DV1}. 
[Since there is exactly one factor of $N^{-1}$ associated with 
each wavevector sum, all of the latter may be changed to integrals,  
using equation (\ref{ident}).]

1. Draw all connected diagrams of $n$ vertices and 
a sink to the left of all vertices; the rightmost vertex must be 
a source.
Each line exiting vertex $j$ must be contracted with a line entering
some vertex $i < j$.  There is a factor
$\delta_{k',-k}$ associated with each such internal
line, where $k$ is the wavevector exiting vertex $j$ and $k'$  
the wavevector entering vertex $i$.  The requirement that all
lines be contracted leads to the condition $2(n_s-1) + n_b - n_c = 0$,
where $n_s$ is the number of sources, $n_b$ the number of bifurcations,  
and $n_c$ the number of junctions.  

2. Each diagram possesses a factor of $(-1)^n$
and a combinatorial factor reflecting the number of
ways of realizing the contractions.  In the series for
$\bar{\rho}$, this factor is given by $2^{C}$, with  
$C=1+n_{3}+2n_{4}+n_s-\ell$, where $n_3$ is the number of 3-vertices 
(of either kind), $n_4$ the number of 4-vertices, 
and $\ell$ the number of simple loops.
  
3. Associated with each bifurcation is a factor 
$2p\omega_{k_{1},k_{2}}=2p[1-\cos k_{1}\cos k_{2}]$. 
Each junction carries a factor $2\omega_{k,0}$ 
and each 4-vertex a factor $\omega_{k_{1},k_{2}}$.  Each source
carries a factor of $p^2 \omega_{k, -k}$.
(The  $k_{i}$ denote the wavevectors exiting the vertex.) 

4. There is a factor $\overline{f}_D$ resulting from the time
integrations, as discussed above.

5. Replace the sink and rightmost source with dressed loops, 
leading to the factor $[4p/(1+4p)]^2$ mentioned above, and 
exclude all diagrams with a 4-vertex in position 1 or $n-1$. 

6. Integrate over all wavevectors.

Collecting the factors of $p$ and $1/p$ associated with the
various vertices, $\overline{f}_D$, and the factor of $p^{-2}$
in the definition of $\overline{\rho}$, we find that
each diagram in the series for $\overline{\rho}$
contains an overall factor $p^{-r}$ where $r = n - n_b - 2(n_s-1)$.
Using the relation $2(n_s-1) + n_b - n_c = 0$, we have 
$r = n - n_c$.  

In order to take advantage of our simple results for the sum of an
infinite set of diagrams represented by the dressed loops, we
adopt $\kappa \equiv (1+4p)^{-1}$ as the expansion parameter
rather than $p$.  Noting that $4p/(1+4p) = 1-\kappa$, and that
$1/p = 4\kappa/(1-\kappa)$, we see that 
the first order diagram (i.e., the single dressed loop of
figure 2) carries a factor of $4/(1+4p) = 4 \kappa$, while
diagrams at higher order carry a factor $[4p/(1+4p)]^2$,
so that at order $1/p^r$ there is an overall factor of
$(4 \kappa)^r/(1-\kappa)^{r-2}$.  Thus diagrams $\propto \kappa^r$
contribute at order $r$ and all higher orders.  Diagrams in this class
must have at least $r+1$ vertices and no more than $3r-2$ vertices.

Enumeration of diagrams at a given order involves (1) identifying 
all allowable sequences of $n$ vertices, and (2) identifying
all possible sets of connections between vertices, for each
sequence.  For diagrams with $n \geq 3$ (i.e., those
not included in the simple dressed loop of figure~2),
vertex 1 (nearest the sink) must be a junction.  (As
explained above it cannot be a 4-vertex.  If it were a bifurcation,
the wavevector of the single line entering this vertex would of
necessity be zero, but such terms have been excluded from
the action.)   For similar reasons, vertex $n-1$ must be either a source
or a bifurcation.  Once the vertex sequence has been fixed, all
possible sets of contractions of outgoing and incoming lines
must be enumerated.  The single line exiting vertex 1 must, naturally,
always terminate at the sink.

The enumeration of sequences and connections is readily codified 
in an algorithm that may be implemented via computer.  In our
routine, for each $n$ and $r$, all sequences (subject to
the above limitations) are constructed.  Then all possible connections
are generated, by simply running through all termination
points for each line independently, and rejecting those sets that
result in uncontracted lines.  In this way we are able to enumerate
all the diagrams at a given order.  

A diagram is specified in terms of
its {\it bond set} $\{(v_1,v'_1),...,(v_m, v'_m)\}$, where $v_j$ and $v'_j > v_j$
are the terminal vertices of line $j$, with $j=0$ for the sink.  Thus the diagram of figure~3  
can be written: (01) (12) (12) (23) (03).
(The computer algorithm was verified against 
hand enumeration up to third order.)
Since the number of diagrams
grows very rapidly, we extended the routine to perform the wavevector
integrations for each diagram generated.  This entails construction
of the numerator and denominator of the integrand, which are
products of factors involving 
the cosines of various linear combinations of wavevectors.
The numerator is a product of factors associated with each vertex,
as noted in item 3 above.  The denominator is a product of factors
associated with each interval between vertices.  These factors
are readily determined, given the vertex sequence and set of
connections.  
Note that there is one free wavevector $k_i$ associated
with each vertex, except for junctions, so that the number of
wavevector sums is $r$.  In the latter case the
wavevector exiting is equal to the sum $K$ of those entering.  
The lines exiting a source carry wavevectors $k'$ and $-k'$,
where $k'$ denotes the new associated wavevector.  In the case of 
a bifurcation or a 4-vertex we may take the wavevectors of the lines
exiting as $k'$ and $K - k'$ where $K$ denotes the wavevector 
entering (or the sum of the wavevectors entering, in the case of a
4-vertex).  Thus we see that the construction of the integrand
(including associated numerical factors) is a straightforward task
that can also be codified in a computational algorithm.
The integrals over the $k_i$ are evaluated numerically using 
a midpoint method \cite{offer}.  Based on results for varying
number of intervals in the numerical integration, we are able to
determine the resulting coefficients with a relative uncertainty of
about $10^{-4}$ or less.

\section{Results}

We have carried out the expansion for $\overline{\rho}$
to order $\kappa^4$.  
Call the number of $n$-vertex diagrams at order $\kappa^r$ $N_{n,r}$ 
and the contribution of this set of diagrams to the coefficient
of $\kappa^r/(1-\kappa)^{r-2}$ in this series $b_{n,r}$; these values
are reported in Table I.

The diagrammatic expansion 
yields 
the following expression for the stationary activity density 
\be
\bar{\rho}_\infty = 1 - \kappa- 1.788 \, 040 \kappa^{2}
 -4.414 \, 481 \kappa^{3} -14.632(2) \kappa^{4} + {\cal O}(\kappa^5)
\label{kappaser}
\ee
In figure 4 we compare equation (\ref{kappaser}) and 
the results of Monte Carlo \cite{RD1} simulations using 
systems of up to 800 sites. 
(For each $p$ value, simulations are performed for various system sizes
and the results extrapolated to the infinite-size limit.)
For $p\geq 3$ the difference between the series expression and simulation 
is less than 0.1\%.  In Ref. \cite{RD1}, a similar degree of precision 
is obtained by extrapolating (using Pad\'e approximants) a 16-term
series (in powers of $t$) to the infinite-time limit.
The present series of four terms appears to furnish (without 
transformation or extrapolation), information equivalent to that
obtained from a much longer series in powers of $t$.
It is worth noting that while the time series is divergent,
the present series appears to be convergent for
small values of $\kappa$, and it is natural to interpret the
first singularity on the positive-$\kappa$ axis as marking the
phase transition.

With a series of only four terms it is of course difficult to draw
firm conclusions regarding the location of the critical point.
We nevertheless
analyze the series via Pad\'e approximants \cite{baker}.  
The [2,2] approximant is the best behaved and is in excellent 
accord with simulation for $p \geq 1.5$.  It yields a critical value of
$p_c = 0.8677(3)$.  (The [1,3] and
[3,1] approximants give $p_c = 0.668$ and 0.702, respectively.)
It is usual to analyse the Pad\'e approximant to the series for the derivative
of the logarithm of the order parameter ($d \ln \overline{\rho}/d \kappa$ in the
present instance), as this function should exhibit a simple pole at the
critical point.  The [2,1] approximant does in fact give an improved estimate
of $p_c = 0.9069$ (about 5\% below the value found in
simulations), while the [1,2] approximant yields $p_c =0.860$.  
(The residue
at the pole of the [2,1] approximant is 0.198, well below any of 
the numerical
estimates for the critical exponent $\beta$ that have been reported,
which suggest $\beta \simeq 0.4$ \cite{DAM,RTO,lubeck}.)
The series prediction is compared against simulation in figure 4.
For $p \geq 3$, series and simulation agree to within uncertainty,
i.e., with a relative error of $10^{-4}$.  We note that the four-term
series for the stationary activity yields results of accuracy
comparable to that obtained from the 16-term series in powers of
time.  The latter, when extrapolated to $t = \infty$, gives
$p_c = 0.906$ \cite{RD1}.

The chief barrier to extending the series is the rapid growth in the
cpu time required in evaluating the multiple integrals over wavevectors,
combined with the explosive growth in the number of diagrams.  (Enumeration
of the diagrams represents a small faction of the computing time.)
Thus the present approach does not appear viable for $r$ greater than
four.

For similar reasons the analysis of the two-dimensional case is
restricted to $r \leq 3$.  As explained in Ref. \cite{DV1},
the formalism remains valid in $d$ dimensions if we replace all
factors $\omega_{k,k'}$ with
\begin{equation}
\omega_{\bf k_1, k_2} = 1 - \lambda_d ({\bf k_1}) \lambda_d ({\bf k_1}) ,
\label{omegadd}
\end{equation}
where
\begin{equation}
\lambda_d ({\bf k}) \equiv \frac{1}{d} \sum_{\alpha = 1}^d \cos k_\alpha .
\label{deflamb}
\end{equation}
Thus $\gamma_k$ in equation (\ref{gamma}) becomes
\be
\gamma_{\bf k}= 4\;p\; [1- \lambda_d ({\bf k}) ] \;.
\label{gammad}
\ee
The expansion involves the same set of diagrams in any dimension; only
the integrals change, with
the wave vectors now ranging over the first Brillouin zone in $d$ dimensions.

In two dimensions our result for the stationary activity density is:

\be
\bar{\rho}_\infty = 1 - \kappa- 1.704 \, 155\kappa^{2}
 -3.7292 \kappa^{3} + {\cal O}(\kappa^4)
\label{kappa2d}
\ee
The series prediction is compared against Monte Carlo simulation in figure 5;
good agreement is observed for $p \geq 1.5$.  The [2,1] and
[1,2] Pad\'e approximants to the three-term series for $\overline{\rho}$
yield critical values of $p_c = 0.507$ and 0.502 respectively,
whereas the estimate from simulation is 0.715.

\section{Correlation functions and probability distribution}

Consider the stationary expectation $\langle n_j n_{j+\ell} \rangle$
of the product of occupations at sites
$j$ and $j+\ell$.  For $\ell \neq 0$ this may be written as
\cite{DV1}:

\be
\label{cell}
C(\ell) \equiv N^{-1} \sum_j \langle n_j n_{j+\ell} \rangle
= N^{-2} \sum_{k} e^{ik\ell} \langle\psi_{k} \psi_{-k} \rangle
\ee        
and separating the $k=0$ term as in equation (\ref{Eq:RHO2}) we find
\be
\label{cella}
C(\ell) = p^2 + 
N^{-2} \sum_{k \neq 0} \cos k\ell \; \langle \psi_k \psi_{-k}
e^{\int_{0}^{t}dt^{'}{\cal L}^{'}_{I}} \rangle_{0}
\ee  
The second term, an infinite sum of diagrams, defines the
connected two-point correlation function $G(|\ell|)$.  The lowest
order contribution comes from the one-vertex diagram (simple loop)
giving

\be
\label{Gell1}
G^{(1)}(|\ell|) = -\frac{p}{4} \int_{-\pi}^{\pi}\frac{dk}{2\pi}
\;\cos k \ell \; (1+ \cos k),
\ee
or $G^{(1)}(1) = - p/8$ and $G^{(1)}(|\ell|) = 0$  for $|\ell| > 1$.  
When the
dressed loop is evaluated this becomes $G^{(1)}(1) = - \kappa p^2 /8$.
The correlation for sites separated by greater than unit distance is
${\cal O} (\kappa^2)$ or higher.  From this result we can draw the following
conclusions: (1) The nearest-neighbor correlation is negative for
large $p$; (2) For large $p$ correlations decay rapidly in space;
(3) As $p \to \infty$, the reduced correlation 
$\overline{G}(\ell) = G(\ell)/p^2$ decays to zero (as $1/p$ or
faster) so that in this limit the site occupancies are
independent random variables.

The stationary expectation of $(\psi_j)^m$ (product of $m$ fields at the
same site) is related to the $m$-th factorial moment of the
one-site occupation distribution.  [For $m=2$ this is seen explicitly in
equation (\ref{Eq:AMG}).]  For $m=3$ for example, we have

\be
\langle n^3 \rangle_F \equiv N^{-1} \sum_j \langle n_j (n_j-1) (n_j-2) \rangle
= N^{-3} \sum_{k,k'} \langle \psi_k \psi_{k'} \psi_{-k-k'} \rangle
\label{fm3}
\ee
which can be written 
\be
\langle n^3 \rangle_F = p^3 
+ 3p N^{-2} \sum_{k \neq 0} \langle \psi_k \psi_{-k} \rangle
+ N^{-3} \sum_{k,k' \neq 0} \langle \psi_k \psi_{k'} \psi_{-k-k'} \rangle
\label{fm3a}
\ee
The second term equals $3 p(\rho - p^2)$ and so is ${\cal O} (p^2)$ for
large $p$.  The third term must be expanded in diagrams in which
the sink has {\it three} lines entering.  The lowest order diagram 
thus involves two vertices, a source and a bifurcation, and
is ${\cal O} (p)$ for large $p$.  We see then that
$\langle n^3 \rangle_F = p^3 [1 + {\cal O} (1/p) ]$ for large $p$.
The same line of reasoning shows that the $m$-th factorial moment 
approaches $p^m$ as $p \to \infty$.  In this limit the one-site
marginal distribution is therefore Poisson with parameter $p$, and
by our previous result on independence,
the joint probability distribution is a product of such distributions.

We defer a detailed analysis of correlation functions to future work,
and stress that the main result of the present section is that in the
large-$p$ limit, the probability distribution is a product of
identical Poisson distributions at each site, as was conjectured
in \cite{RD1}.  It is readily seen that this remains valid in
$d \geq 2$ dimensions.

\section{Discussion}

We derive a path-integral representation and diagrammatic expansion
for the stationary activity density in a stochastic sandpile with a
conserved particle density.  Because of conservation, the $k=0$
Fourier mode of the particle density (and associated field $\psi_k$)
has a fixed value, rather than being a random variable.  This observation
permits us to reorganize the effective action so that the control
parameter $p$ appears explicitly, rather than being defined implicitly
in the initial condition.  The bilinear part of the action now
describes diffusion at a rate $4p$.  Because of this, the propagator
carries an exponential factor, and all time integrations can be
realized to obtain the limiting ($t \to \infty$) activity directly.
The ensuing expansion for $\overline{\rho}_\infty$ involves the
parameter $\kappa = (1+4p)^{-1}$, i.e., this is a large-$p$ expansion.
(As noted in Ref. \cite{RD1}, the time series is also most useful
for large $p$ values.)  We are able to sum certain infinite classes
of diagrams through the device of ``dressed loops."  Despite this,
the number of diagrams to be evaluated at each order grows
explosively, so that our final calculational result (the activity series
to ${\cal O}(\kappa^4)$ is quite modest.  
The fourth-order series agrees very well with simulation in the
supercritical regime, and yields (via Pad\'e approximation)
the critical value $p_c$ to within about 10\%.  
A similar favorable comparison is seen in the 
two-dimensional case, although the three-term series furnishes a poorer
estimate for $p_c$.
Given these encouraging results, it is reasonable
to hope that extended series will yield quantitative predictions for 
critical properties.
We have also used the reorganized expansion to show that in the
large-$p$ limit, the sandpile is governed by Poisson-product distribution.
Our results strengthen
the conclusion, until now based on simulation and mean-field-like
analyses, that fixed-energy sandpiles exhibit a phase transition
as the particle density is varied.  It is of great interest to
know if the details of this transition can be analysed using the
operator and path-integral formalisms.
\vspace{2em}

{\small {\bf ACKNOWLEDGMENTS}}
\vspace{1em}

We are grateful to Miguel A. Mu\~noz for valuable comments on the
manuscript.
This work was supported by CNPq and CAPES, Brazil.

\newpage

\begin{table} 
\begin{center} 
\begin{tabular}{|r|r|r|r|} 
$n$        &  $r$  & $N_{n,r}$ & $b_{n,r}$ \\         
\hline 
3          &   2   &  2  & $-2.384 \, 052$        \\  
4          &   2   &  3  & $ 0.596 \, 013$         \\  
\hline   
4          &   3   &   4  & $  5.625 \, 989 $      \\  
5          &   3   &  49  & $-19.520 \, 916 $      \\  
6          &   3   & 180  & $ 11.376 \, 647 $      \\  
7          &   3   & 306  & $ -1.896 \, 110 $      \\  
\hline   
5          &   4   &          8   & $ -13.225 \, 188 $   \\  
6          &   4   &        311   & $  135.895\, 511  $  \\  
7          &   4   & $3 \, 471 $  & $ -311.353 $         \\  
8          &   4   & $21 \, 961$  & $  256.075 $         \\  
9          &   4   & $76 \, 261$  & $  -88.685 $         \\  
10         &   4   & $136 \, 404$ & $   11.092 $   \\  
\end{tabular} 
\end{center} 
\label{tab1} 
\noindent{\sf Table I. Number of diagrams $N_{n,r}$ and coefficient $b_{n,r}$
 in the expansion of the activity.}
\end{table}

\newpage

\noindent{\bf FIGURE CAPTIONS}
\vspace{1em}

\noindent Figure 1.  Vertices (a - d) in the interaction ${\cal L}'$ and the sink (e)
representing the activity density.
\vspace{1em}

\noindent Figure 2.  Definition of a ``dressed loop" as the sum of one, two, three,...
simple loops joined at 4-vertices.
\vspace{1em}

\noindent Figure 3.  A three-vertex diagram. 
\vspace{1em}

\noindent Figure 4. Scaled stationary activity density $\overline{\rho}$ versus particle
density $p$ in one dimension.  Upper curve: series prediction, equation (\ref{kappaser}); 
the curve labeled [2,1] is obtained by integrating the Pad\'e approximant
to the series for $d \ln \bar{\rho}d \kappa$;
points: Monte Carlo simulation.  Error bars are smaller than the symbols.  
\vspace{1em}

\noindent Figure 5. Scaled stationary activity density $\overline{\rho}$ versus particle
density $p$ in two dimensions.  Upper curve: series prediction, equation (\ref{kappa2d}); 
lower curve: [2,1] Pad\'e approximant to the series;
points: Monte Carlo simulation.


\begin{thebibliography}{dillo 83}
\bibitem{offer} The enumeration/integration code is available on request
to the authors.
\end{thebibliography}
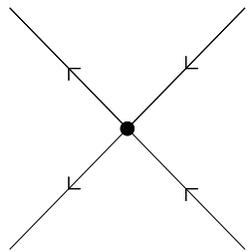
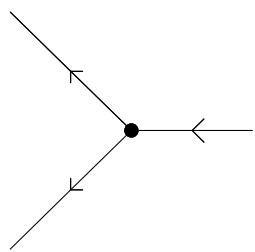
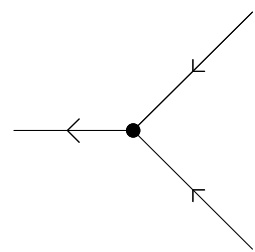
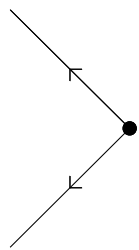
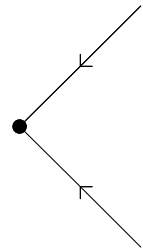
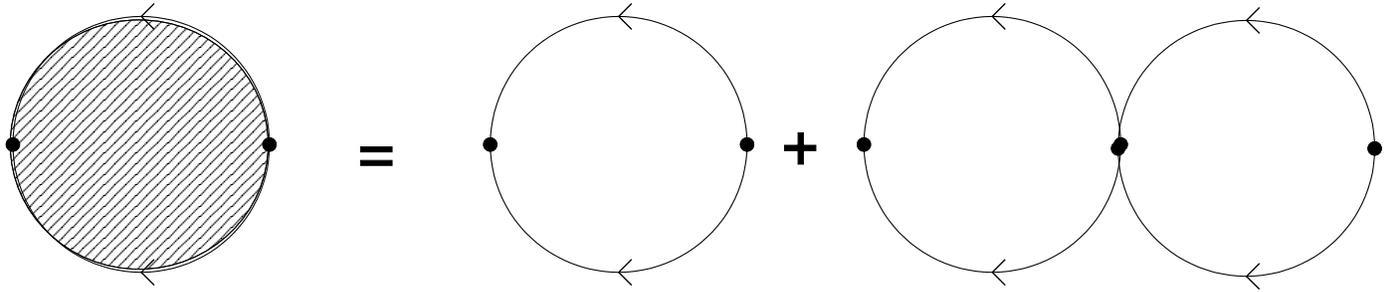
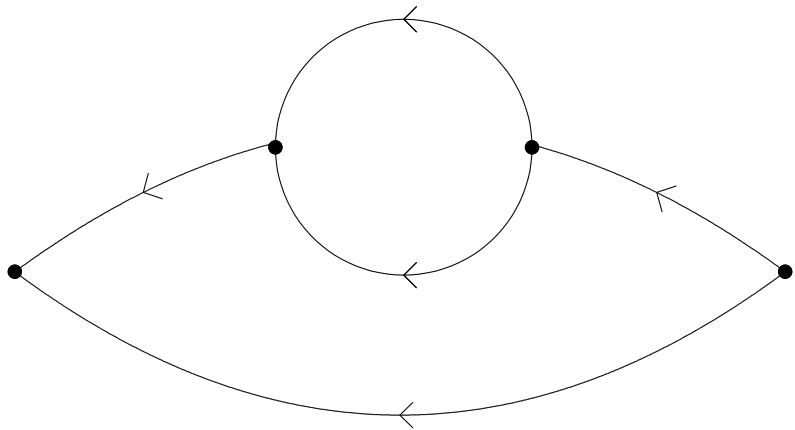
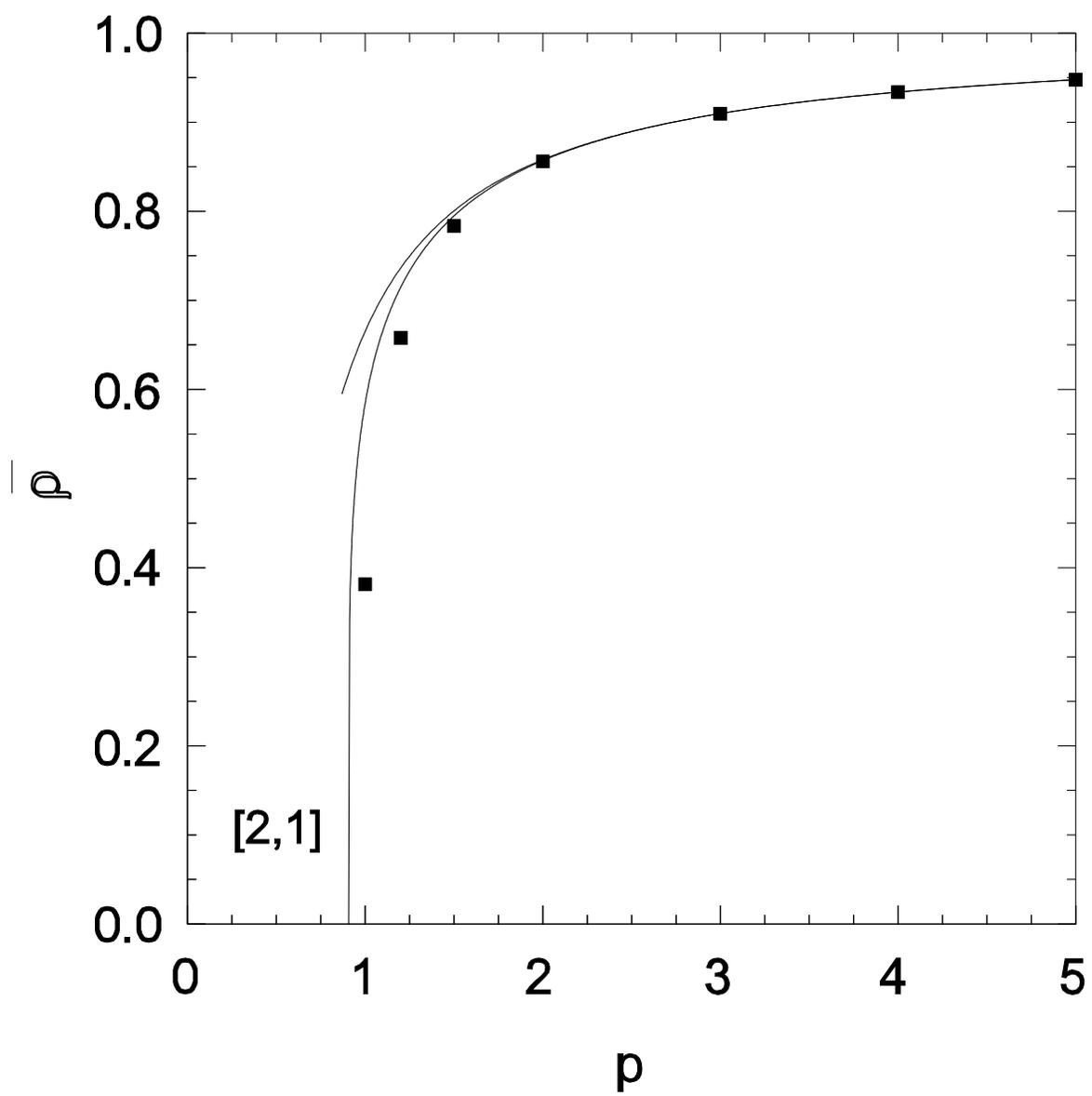
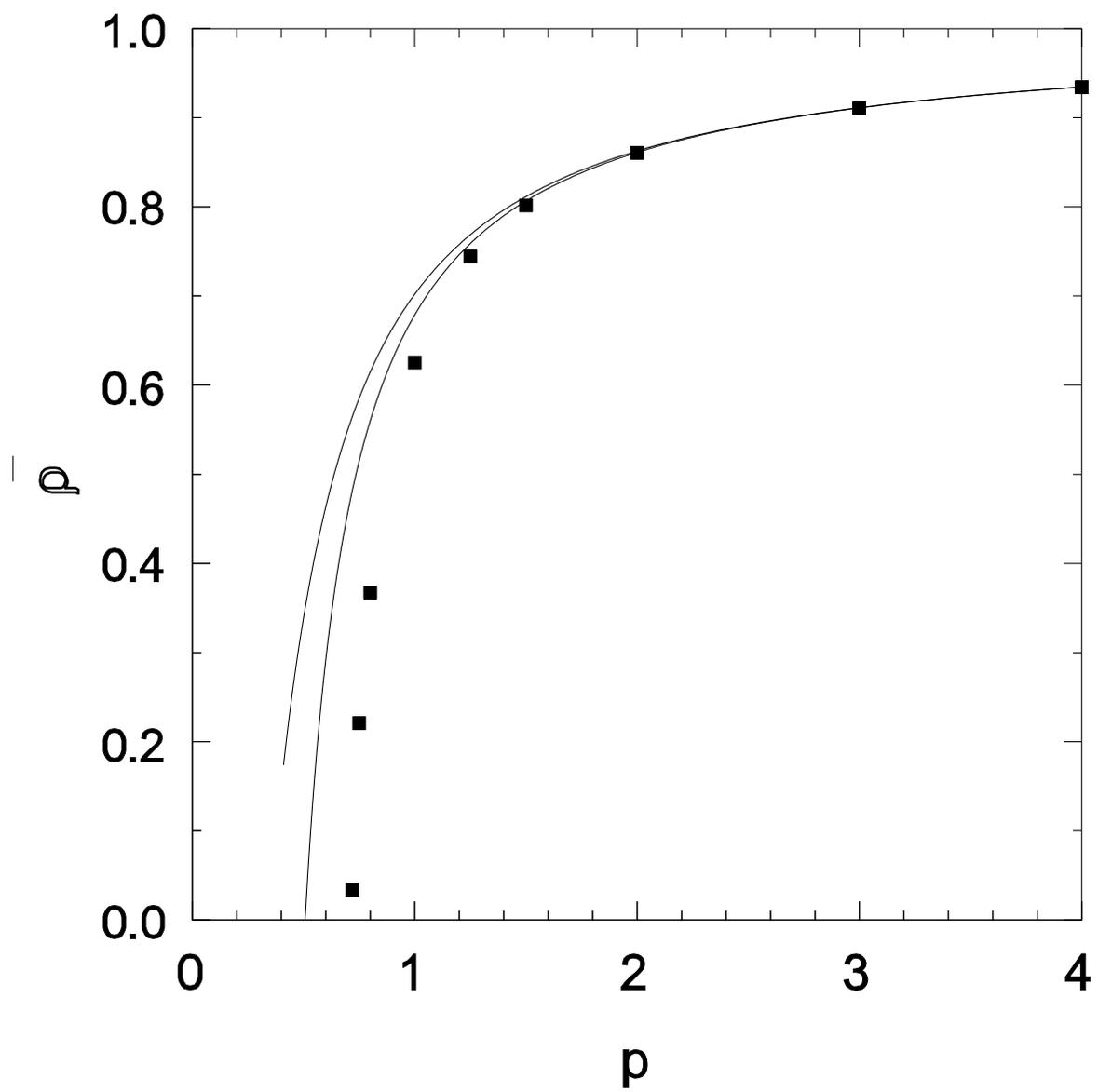
\end{document}